\documentclass[12pt]{iopart} 

\usepackage{epsfig}
\usepackage{graphicx}
\usepackage{dcolumn}
\usepackage{bm}

\usepackage[ansinew]{inputenc}
\usepackage[T1]{fontenc}
\usepackage{textcomp,pstricks}

\usepackage{SIunits}
\usepackage{color}

\begin{document}
\title{Tomographic readout of an opto-mechanical interferometer}

\author{Henning Kaufer$^{1}$, Andreas Sawadsky$^{1}$, Tobias Westphal$^{1}$, Daniel Friedrich$^{1,\footnotemark[3]}$, and Roman Schnabel$^{1,*}$}
\address{
$^{1}$ Institut f\"ur Gravitationsphysik, Leibniz Universit\"at Hannover and\\ Max-Planck Institut f\"ur Gravitationsphysik (Albert-Einstein-Institut), 30167 Hannover, Germany\\
$^{*}$ Corresponding author: roman.schnabel@aei.mpg.de}

\begin{abstract}
The quantum state of light changes its nature when being reflected off a mechanical oscillator due to the latter's susceptibility to radiation pressure. As a result, a coherent state can transform into a squeezed state and can get entangled with the motion of the oscillator. The complete tomographic reconstruction of the state of light requires the ability to readout arbitrary quadratures. Here we demonstrate such a readout by applying a balanced homodyne detector to an interferometric position measurement of a thermally excited high-Q silicon nitride membrane in a Michelson-Sagnac interferometer. A readout noise of $\unit{1.9 \cdot 10^{-16}}{\metre/\sqrt{\hertz}}$ around the membrane's fundamental oscillation mode at $\unit{133}{\kilo\hertz}$ has been achieved, going below the peak value of the standard quantum limit by a factor of $8.2$ (9 dB). The readout noise was entirely dominated by shot noise in a rather broad frequency range around the mechanical resonance.
\end{abstract}

\footnotetext[3]{Present address: Institute for Cosmic Ray Research, The University of Tokyo, 5-1-5 Kashiwa-no-Ha, Kashiwa, Chiba 277-8582, Japan}

\section{\label{sec:level1}Introduction}
Experiments in quantum opto-mechanics seek to create and study pure quantum states in compound systems of mechanical oscillators and light. Recently, the quantum mechanical ground state of a mechanical oscillator was reached using laser cooling based on the light's radiation pressure \cite{PainterGndState}. Radiation pressure coupling of light and mechanical devices also forms the basis for proposals to produce (ponderomotively) squeezed states of light \cite{PondSqueez} as well as to produce entanglement between light fields and mechanical motion \cite{VGFBTGVZA07,entanglementboseeinstein} or between two mechanical objects \cite{Helge08} providing a means for reliable storage of quantum states \cite{quantummemory}. 
A tomographic characterization of the optical and mechanical subsystems involved are necessary to provide a full description of the quantum state.

Here we apply a broadband tomographic characterization of the optical subsystem to an opto-mechanical system around its mechanical resonance, i.e. a tomographic interferometer readout. The system is a Michelson-Sagnac interferometer containing a translucent, high-Q mechanical oscillator. The readout scheme is a balanced homodyne detector (BHD) allowing the measurement of arbitrary field quadratures, which is necessary for a generic full description of the state. 

\begin{figure}[!t]
\begin{center}
\includegraphics{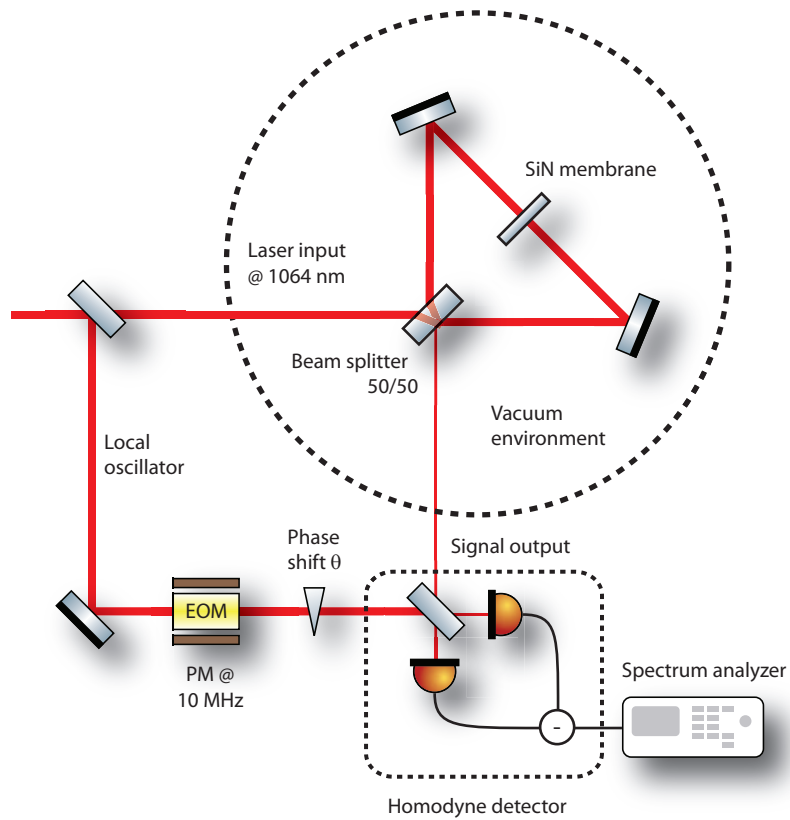}
\end{center}
\caption{\label{fig:ifohomo} Sketch of the Michelson-Sagnac interferometer with homodyne readout scheme. The laser light is split by a balanced beam splitter and recombined after reflection at the membrane. Signal sidebands leak out the dark port and are overlapped with a strong local oscillator (LO) on a second (homodyne) beam splitter. An electro-optical modulator (EOM) provided a phase modulation (PM) to generate an error signal for locking the homodyne readout phase $\theta$.}
\end{figure}

In figure (\ref{fig:ifohomo}) the Michelson-Sagnac interferometer of our experiment is sketched within the dashed circle. It consisted of a balanced (50/50) beam splitter and two steering mirrors forming its Sagnac mode. A commercially available silicon nitride (SiN) membrane  \cite{harris2} was placed in the center of the folded arms. The reflected parts of the light were overlapped with the respective transmitted parts and formed the interferometer's Michelson mode, which was sensitive to a displacement of the membrane. The interference of all four beams  at the beam splitter towards the interferometer's signal output was controlled to be destructive in order to provide a so-called 'dark port'. A detailed description can be found in \cite{Friedrich1}. The topology's quantum noise contributions have been theoretically analyzed in \cite{Kazuhiro} and the first experimental realization of the topology was reported in \cite{Friedrich1,FuPaper} using a conventional single photodiode readout. Recently, the Michelson-Sagnac topology was theoretically investigated in view of the realization of dissipative opto-mechanical coupling \cite{xuereb2011}. 

The membrane used in this work had a measured power reflectivity of $ R = 17\,\%$ at a laser wavelength of $\unit{1064}{\nano\metre}$ under normal incidence. By taking into account the material's index of refraction $n_{\textnormal{SiN}} = 2.2$ \cite{harris3} its thickness could be deduced to $\unit{40}{\nano\metre}$. This together with the membranes side length of $\unit{1.5}{\milli\metre}$ results in an effective mass of $m = \unit{80}{\nano\gram}$ \cite{FuPaper}. The interferometer was operated inside a vacuum chamber to avoid gas damping or excitation of the membrane motion. We determined the mechanical quality factor $Q$ of the fundamental oscillation mode for different pressures. As result, we found $Q = 6 \cdot 10^{5}$ for gas pressures below $4 \cdot \unit{10^{-6}}{\milli\bbar}$.

A piezoelectric element (piezo) actuated the membrane's position such that the carrier light destructively interfered in the interferometer's signal output port. The signal field thus mainly contained the upper and lower sideband fields that were produced by the thermally excited membrane motion. To accomplish the tomographic readout, the output field was overlapped with an external local oscillator (LO) field of the same optical frequency as the interferometer's input on a 50/50 beam splitter. The light of the two beam splitter output ports was directed to two photodiodes forming a balanced homodyne detector. The difference of the photocurrents depends on the relative phase $\theta$ between the LO and signal field and can be written as
\begin{equation}
i_{-}(\theta) \propto \alpha_{\rm{LO}}  X_{\theta} = \alpha_{\rm{LO}} (\cos(\theta) X_1 + \sin(\theta)X_2) \,.
\label{homosub}
\end{equation}

Here, $\alpha_{\rm{LO}}$ is  the coherent amplitude of the local oscillator, $X_1$ and $X_2$ are the amplitude and phase quadrature amplitudes of the signal field with respect to its coherent amplitude, and $X_{\theta}$ is the quadrature amplitude at a phase angle $\theta$. Note, that the above equation is only valid if $\alpha_{\rm{LO}}$ is much bigger than the coherent amplitude of the signal beam, which is close to zero (exactly zero in case of perfect interferometer contrast) for the interferometer operated on its dark port.

\begin{figure}[!t]
\begin{center}
\includegraphics{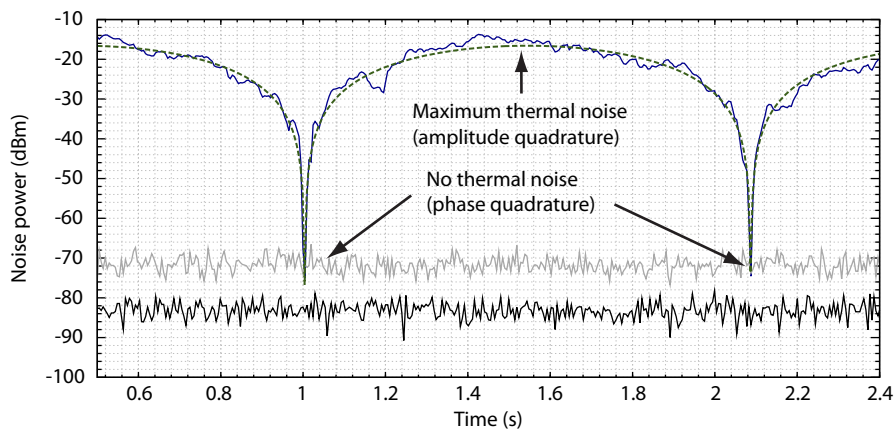}
\end{center}
\caption{\label{fig:scanhomophase} Zero-span noise-power measurement at the membrane's fundamental resonance frequency $f_{\textnormal{res}} = \unit{133.88}{\kilo\hertz}$. The readout phase $\theta$ is scanned continuously. The blue curve shows the measured noise power, the dashed green line its theoretical model, and the grey curve shows the independently measured shot noise. For a particular quadrature phase ($\theta=\pi/2$), no membrane displacement but only shot noise was measured. The black curve shows electronic dark noise. The resolution bandwidth (RBW) was set to $\unit{10}{\kilo\hertz}$.}
\end{figure}

The membrane's thermally driven motion generates an excitation of the reflected light's phase quadrature, which is converted to an amplitude quadrature excitation of the interferometer's output field by the interference at the beam splitter. Vice versa, the output field's phase quadrature is a measure of the differential amplitude quadrature excitation in the two arms of the interferometer. Due to the strong laser noise rejection at the Michelson dark-port this excitation is given by optical quantum noise, i.e. by shot noise.

\begin{figure}[!t]
\begin{center}
\includegraphics{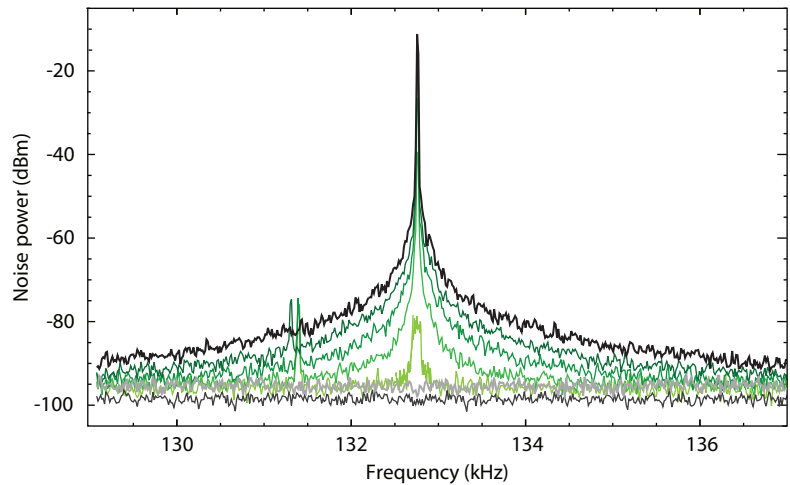}
\end{center}
\caption{\label{fig:spectra} Noise power spectra taken for different values of the quadrature angle $\theta$. The top curve was taken with $\theta=0$ (black line) while the horizontal trace corresponds to shot noise (gray line). As $\theta$ approaches $\pi/2$, less information about the membrane displacement is captured. Electronic dark noise is plotted in dark-gray.
}
\end{figure}

\begin{figure}[!b]
\begin{center}
\includegraphics[width=12cm]{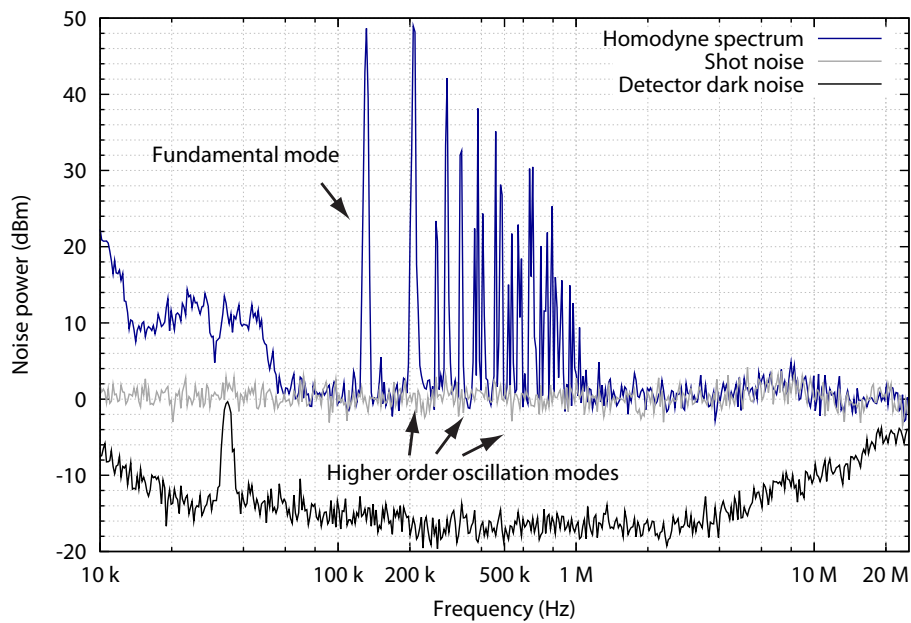}
\end{center}
\caption{\label{fig:breitbandig} Noise power spectrum taken while $\theta = 0$. Above $\unit{50}{\kilo\hertz}$ the spectrum is limited by optical shot noise. For smaller frequencies, accoustics and other noise sources are present. The upper readout limit of $\unit{25}{\mega\hertz}$ is set by electronics only. The RBW was $\unit{10}{\kilo\hertz}$ and the data has been normalized by the modelled electronic transfer function of the photo detector. The bump at $\unit{8}{\mega\hertz}$ is due to imperfect modelling.}
\end{figure}

Figure (\ref{fig:scanhomophase}) shows a zero-span measurement of the noise power at the membrane's resonance frequency $f_{\textnormal{res}}$ while the readout quadrature angle $\theta$ was scanned by a piezo driven mirror. This measurement is a tomographic readout because it allows for reconstruction of the output light's Wigner-function. 
Indeed the motion of the membrane is not visible in one particular quadrature angle where the noise power reaches shot noise. Note that squeezing is not expected in our setup due to the low laser power and the high level of thermal noise.
In figure (\ref{fig:spectra}) measured spectra of the output light around the membrane's eigenfrequency are shown for various guadrature angles. For a readout quadrature of $\theta = \pi/2$ almost no signal from the membrane excitation is visible. In figure (\ref{fig:breitbandig}) we present the  broadband power spectrum for a frequency range from $\unit{10}{\kilo\hertz}$ to $\unit{25}{\mega\hertz}$. Here, we stabilized the readout homodyne phase to the amplitude quadrature $\theta = 0$. For this purpose, an electro-optic modulator (EOM) imprinted a phase modulation of $f_{\textnormal{mod}}=\unit{10}{\mega\hertz}$ on the local oscillator beam. The homodyne signal was demodulated with the modulation frequency, generating an error signal for locking the homodyne readout phase via the above mentioned piezo driven mirror. The read out is limited by quantum shot noise for frequencies above $\unit{50}{\kilo\hertz}$.

\begin{figure}[!t]
\begin{center}
\includegraphics{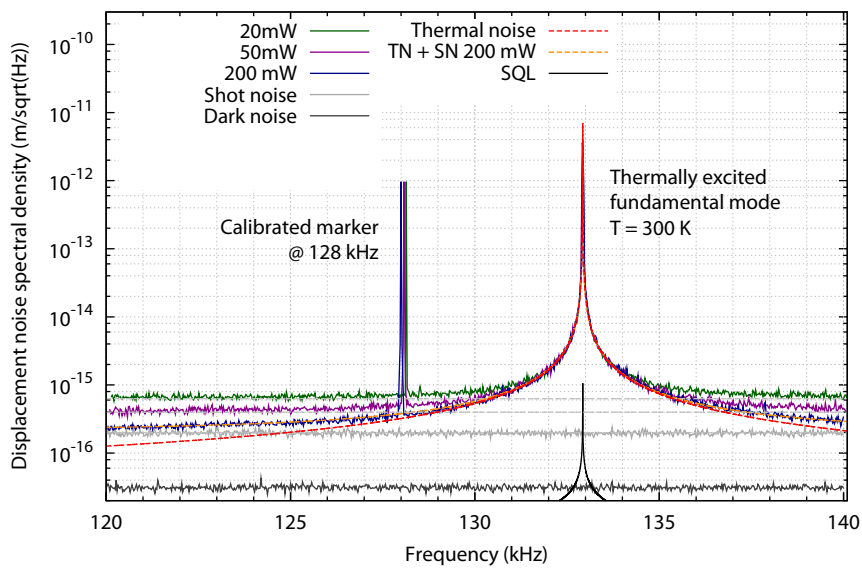}
\end{center}
\caption{\label{fig:powers} Calibrated spectra for different input powers $P_{\textnormal{in}}$. The measured shot noise level for $P_{\textnormal{in}}=\unit{200}{\milli\watt}$ is plotted in gray. The corresponding electronic dark noise level is plotted as brown line. Each dashed gray line indicates the calculated shot noise level for each individual input powers. The orange dashed line shows the uncorrelated sum of shot noise and thermal noise for $\unit{200}{\milli\watt}$ input power. It is in excellent agreement with the measurement. The standard quantum limit is plotted as solid black line. A defined displacement marker at $\unit{128}{\kilo\hertz}$ is used for calibration of the $y$-axis.}

\end{figure}

In figure (\ref{fig:powers}) we characterized the sensitivity of the Michelson-Sagnac interferometer with balanced homodyne detector readout for three different interferometer light powers.
The spectra were converted into displacement spectral densities by the method described in \cite{FuPaper} and independently by a calibrated piezo driven membrane motion with $f=\unit{128}{\kilo\hertz}$.  
Starting from $P_{\textnormal{in}} = \unit{20}{\milli\watt}$ we increased the input power to $\unit{200}{\milli\watt}$.  The phase of the balanced homodyne detector's local oscillator was controlled to the amplitude quadrature ($\theta =0$) and its power was $P_{\textnormal{LO}} = \unit{12}{\milli\watt}$ for all measurements. The signal power in all measurements was always less than $\unit{0.5}{\milli\watt}$.

The quantum shot noise was measured by blocking the interferometer output at the BHD. A comparison between the calibrated shot noise and its theoretical prediction (1) carried out in \cite{FuPaper} indicated an overall loss of $50\,\%$ in the experiment. This value agreed with our expectations and was given by the imperfect quantum efficiency of our balanced homodyne detector of separately mesasured $70\,\%$ and imperfect reflectivities of the interferometer input and output optics. We expect that with better photo diodes and high-quality optics the overall loss can be reduced to below $10\,\%$ in future experiments. For each input power, the optical shot noise is plotted in grey (dashed lines). Off resonance, it is the dominant noise source within a broad frequency range from $\unit{50}{\kilo\hertz}$ to $\unit{25}{\mega\hertz}$ in all measurements.
For $P_{\textnormal{in}}=\unit{200}{\milli\watt}$ a displacement sensitivity of $\unit{1.9\cdot 10^{-16}}{\metre/\sqrt{\hertz}}$ at $\unit{120}{\kilo\hertz}$ could be achieved. The orange dashed line corresponds to the uncorrelated sum of the thermal and shot noise. The gap between the three different measured shot noise levels exactly matches $\sqrt{10}$ and $\sqrt{4}$ as predicted by theory.

To conclude, we applied a tomographic readout to fully characterize the state of light being reflected off a thermally excited SiN membrane at room temperature. The readout was shot noise limited over a broad spectral region around the mechanical resonance and achieved an imprecision significantly below the standard quantum limit at the membrane's resonance frequency. 
In comparison to a simple amplitude quadrature readout using a single photodiode, the balanced homodyne detector has the advantage that the interferometer can be operated precisely at a dark port providing full laser noise rejection. Additionally, the readout local oscillator acts as a mode selective element, which is useful to discriminate residual transversal modes at the interferometer signal output port. Our setup is in principle able to detect nonclassical properties of the output light such as ponderomotive squeezing \cite{PondSqueez} or could be part of an opto-mechanical entanglement analysis \cite{VGFBTGVZA07}. As the displacement sensititivity is limited by quantum shot noise, it could be enhanced by injecting squeezed states of light as demonstrated in numerous table-top experiments \cite{Kimble,McKenzie,Eberle}. 
 
\ack
We acknowledge fruitful discussions with Y. Chen, S. L. Danilishin, K. Danzmann, K. Hammerer, F. Ya. Khalili and H. Miao. This work has been supported by the International Max Planck Research School for Gravitational Wave Astronomy (IMPRS), and by the Centre for Quantum Engineering and Space-Time Research (QUEST). H.K. acknowledges the HALOSTAR scholarship program.

\section*{References}

\bibliographystyle{ol}


\end{document}